\documentclass{elsarticle}

\usepackage{lineno,hyperref}
\modulolinenumbers[5]

\journal{Journal of \LaTeX\ Templates}

\usepackage{numcompress}\biboptions{authoryear}

\begin{document}

\begin{frontmatter}

\title{Multi-wavelength view of an M2.2 Solar Flare on 26 November 2000}

\author{R. Chandra}
\address{Department of Physics, DSB Campus, Kumaun University, Nainital -263 001, India}
\cortext[mycorrespondingauthor]{Corresponding author}
\ead{rchandra.ntl@gmail.com}
\author{V. K. Verma}
\address{Uttarakhand Space Application Center, Dehradun -- 248 006, India}
\author{S. Rani}
\address{Department of Physics, DSB Campus, Kumaun University, Nainital -- 263 001, India}
\author{R. A. Maurya}
\address{National Institute of Technology, Calicut -- 673 601, India}

\begin{abstract}
In this paper, we present a study of an M2.2 class solar flare of 26 November 2000
from NOAA AR 9236. The flare was well observed by various ground based
observatories (ARIES, Learmonths Solar Observatory) and space borne instruments (SOHO, HXRS, GOES) in
time interval between 02:30 UT to 04:00 UT. The flare started with long arc-shape
outer flare ribbon. Afterwards the main flare starts with two main ribbons. Initially
the outer ribbons start to expand with an average speed ($\sim$ 20 km s$^{-1}$) and
later it shows contraction. The flare was associated with partial halo coronal mass ejection
(CMEs) which has average speed of 495 km s$^{-1}$. The SOHO/MDI observations show that the active region was in 
quadrupolar magnetic configuration. The flux cancellation was observed before the flare onset close to flare site. 
Our analysis indicate the flare was initiated by the magnetic breakout mechanism.
\end{abstract}

\begin{keyword}
\texttt{Sun: solar flares - Sun: reconnections - Sun: coronal mass ejections}
\end{keyword}

\end{frontmatter}

\section{Introduction}

Solar flare are large explosion in the Sun's atmosphere that can release as much as
6$\times 10^{25}$ joules of energy (for review see \citet{benz08}). Solar flares affect all layers of the
solar atmosphere, heating plasma to tens of millions of Kelvin and accelerating 
electrons, protons \citep{Gopalswamy04,Chandra13}.
They produce radiation across the whole electromagnetic
spectrum i.e. from radio waves to gamma rays. Major goals of solar
flare research is to determine the origin and evolution of the energetic electrons
accelerated during impulsive phase of the solar flares. These particles are most
directly observable through their gyro-synchrotron radiations at microwave (MW)
frequencies and emission of hard X-ray (HXR) through collisional bremsstrahlung
\citep{Holman84}.

It is now widely accepted that solar flares are produced by the release of
energy stored in the stressed magnetic field. Magnetic reconnection is mainly responsible for 
this energy release. Continuous flux emergence in
the active region is one of the main indicator for the stressed magnetic field,
whereas the flux cancellation can tell about the reconnection. 
The magnetic flux  may already be twisted while emerging in the
photosphere. The photospheric motion of magnetic polarities are the drivers of the free
energy storage and the emerging flux would provide the trigger mechanism for the
impulsive energy release. Therefore magnetic flux emergence and cancellation are
very important to understand the causes of flare initiation. 

The release energy during solar flares produces accelerated electrons and ions, which interact with the
ambient solar atmosphere. High temporal resolution observations are very important to investigate the time 
evolution of solar flares. This allow us to study the different phases of flare evolution viz. impulsive, main 
and gradual phase. Therefore, analysis of multi-wavelength data with high spatial and temporal resolution is 
crucial for understanding the acceleration and propagation of particles during solar flares 
(for example see \cite{Aulanier00, Galsgaard00,Fletcher01, Chandra06, Chandra13} and references therein).

It is now established that flares and coronal mass ejections (CMEs)
are predominantly magnetic explosions and they show rapid motion and heating
that  infer to be driven by magnetic energy locally contained in the magnetic
field \citep{Svestka76, Sturrock80}. All CMEs and many flares exhibit
outward mass motion, even though it is likely that part of the magnetic field must
shrink (implode) in order that there be an overall decrease in magnetic energy in
the region of the explosion \citep{hudson00}.  The origins of CMEs were earlier investigated by 
\citet{Verma89} and \citet{Verma92,Verma02} and  suggested that the CME events 
are perhaps produced by some mechanism, in which the mass ejected by 
some solar flares or active prominences, gets connected with the open magnetic 
lines of CHs (coronal holes: source of high speed solar wind streams) and moves   
along them to appear as CMEs.

To explain the different observational features of solar flares and associated phenomena, the
CSHKP model was proposed \citep{Carmichael64,Sturrock66,Hirayama74,Kopp76}. This model explains the 
filament eruptions, flare ribbon formation and their separation.  
The CSHKP model is two-dimensional. Recently 
\citet{Aulanier13} and \citet{Janvier13} extend the 2D CSHKP
model into 3D using MHD simulations and compare the numerical
results with high resolution {\em Solar Dynamics Observatory} ({\it SDO})
observations. However, these models are not to explain the trigger mechanism of solar eruptions.

To understand the trigger mechanism of solar eruptions several models have been put forward. 
These models includes: Magnetic breakout \citep{Antiochos99}, Tether 
cutting \citep{Moore01}, kink instability \citep{Torok04}, ideal MHD instability 
(torus instability) \citep{Forbes91,Kliem06}. However, it is still not settled which model 
initiates the solar eruptions. Therefore, to understand the solar eruptions trigger mechanism 
is still major problem in solar physics. 

Here, we discuss about the ``magnetic breakout'' model, because it appears to be favorable 
mechanism for our studied event. According to this model, the eruption is 
triggered at the magnetic null--point high in the corona in a quadrupolar magnetic configuration.
If the magnetic arcade starts to rise due to shear in magnetic field, its expansion results in the
formation of a current sheet and reconnection at null--point. Due to the reconnection, the magnetic tension 
decreased continuously and as a result the eruption starts. 

Several studies have been done on the observational evidence of magnetic breakout 
model (\citet{Aulanier00, Mandrini06, Joshi07, Chandra11, Aurass11, Reeves15, chen16} and references cited therein). 
Let us discuss these studies in brief. \citet{Aulanier00} did the detailed analysis of Bastille day flare of 14 July, 1998 and concluded the eruption was due to the magnetic breakout. With the help of magnetic field extrapolations, they found the presence of a null point in the corona. \citet{Mandrini06} studied the precursor phase of X17 flare on 28 October, 2003. They have 
reported a precursor event in  H$\alpha$ and TRACE images in a  large-scale quadrupolar 
reconnection and concluded that this precursor contributes to decrease 
magnetic tension and allows the filament to erupt in a way similar to breakout model. 
However, they showed that the magnetic reconnection occurs in Quasi-Separatrix Layers (QSLs) 
instead of null point. \citet{Chandra11} also presented the evidence of magnetic breakout mechanism 
in the flare of 20 November, 2003 without the null-point topology. \citet{Joshi07} and \citet{Aurass11}  
shows the evidence of magnetic breakout on the basis of radio  observations.
Very recently \citet{chen16} presented the observations of imaging of magnetic breakout using the high 
resolution SDO data and found the breakout reconnection sets around 40 min before the main flare onset.
From the above discussion on  magnetic breakout mechanism, it is still
not clear when and where the eruption is triggered in this mechanism. Hence, it needs further 
investigation.

Therefore to shed light on the above discussed problems, in this present paper, we have investigated 
an M2.2 solar flare and CMEs observed on 26 November, 2000 from NOAA AR 9236 in view of
multi-wavelength observations. In section 2, we mentioned
about observational data sources and an overview of active region.
The results are presented in Section 3 and conclusion is given in section 4.

\section{Observations and overview of the active region NOAA 9236}

The active region, NOAA AR 9236 was one of the most flare productive active
region in the peak phase of solar cycle 23. 
The active region appeared in the east limb on 18 November, 2000 as $\beta$ magnetic 
configuration and went over the west limb on 30 November, 2000. 
The magnetic configuration of active region becomes $\beta\gamma$ on 23 November, 2000. 
During the disk passage the active region produced 40 C--class, 06 M--class, 10 X--class GOES X--ray 
flares. Many flares from this active region were studied by several 
authors \citep{Nitta01, wang02, moon03, takashi04}. In these studies, it was reported that the 
active region was characterized by continuous magnetic flux emergence in the penumbral area sunspot. 
According to \cite{wang02}, the flux emerged in the active region between 23 to 25 November, 2000. 
Afterwards the flux emergence stops. 

On 2000 November 26, we observed an M2.2 class GOES flare /optical class 1F flare 
in H$\alpha$ from ARIES, Nainital in this active region. The active region was located on 
solar disk at N22W34 on that day. The flare impulsive phase starts 
around 02:47 UT, peaked at 03:08 UT, and ended around 03:20 UT. 

In order to know the magnetic causes of the flare, we have presented the magnetic filed images before the 
flare onset in Figure \ref{magnetic}. In the figure the top image shows the extended view of active region. 
The active region is in quadrupolar magnetic configuration. The negative/positive polarities of the quadrupole is shown by
N1N2/P1P2 and enclosed by white and black contours respectively. The enlarged view of the  active region shown 
by square in the top image is shown in the middle and bottom panel of the figure. We have circled the area of flare location. Looking at these circled regions, we  noticed the significant flux cancellation before the onset of M2.2 
class flare. 

The observational data used in this study is taken from the following instruments:

\begin{itemize}
\item{\bf {H$\alpha$ Data:}}
For the current study we used the H$\alpha$ data observed from ARIES (formerly State Observatory), Nainital, India 
with 15 cm f/15 Coud\'e--refractor telescope. The telescope was equipped with high
speed CCD camera developed for observations of solar flares, which is capable to
record flare image at time interval of 25 ms. Details of the CCD camera system
is described by \cite{Verma99}. The resolution of images are 0.65 arcsec. The cadence of images 
during the observations was from one to ten sec during the flare observation.

\item{{\bf X-ray and Radio Data:}}
To understand the thermal and non-thermal nature of the flare, we use the X-
ray data from the Czech-made Hard X-Ray Spectrometer (HXRS) instrument on
board the Multispectral Thermal Imager (MTI) satellite \citep{farnik01}. In
addition of these data, we have used the radio data of different frequencies
observed from Learmonth Solar Observatory.

\item{{\bf SOHO/MDI and SOHO/EIT  Data:}}
In order to study the magnetic complexity of the active region, we have used the data from  Michelson Doppler Imager (MDI) \citep{sche1995} onboard Solar and Heliospheric Observatory (SOHO) satellite. 
The cadence and the pixel size of the images are 96 min and 1.98 arcsec respectively. We have also
used the SOHO/EIT  Fe XII (195 \AA) data \citep{delab1995}. The cadence of EIT data 
was 12 min, and the pixel resolution was 2.5 arcsec.

\item{{\bf LASCO CME  Data:}}
For the associated CME with the studied flare, we used the data from Large Angle Spectroscopic Coronagraph (LASCO; \cite{Brueckner95}) C2 data. 

\end{itemize}

\section{Analysis and Results}

To understand the multi-wavelength spatial and temporal characteristics of the flare, in 
this section, the evolution of the flare observed in H$\alpha$, radio, X-ray and EUV wavebands is presented.

\subsection{Spatial Evolutions}

\begin{figure}[t]
\vspace*{-2.7cm}
\centering
\hspace*{-2cm}
\includegraphics[width=0.9\textwidth,viewport=  80 190 480 800, clip=]{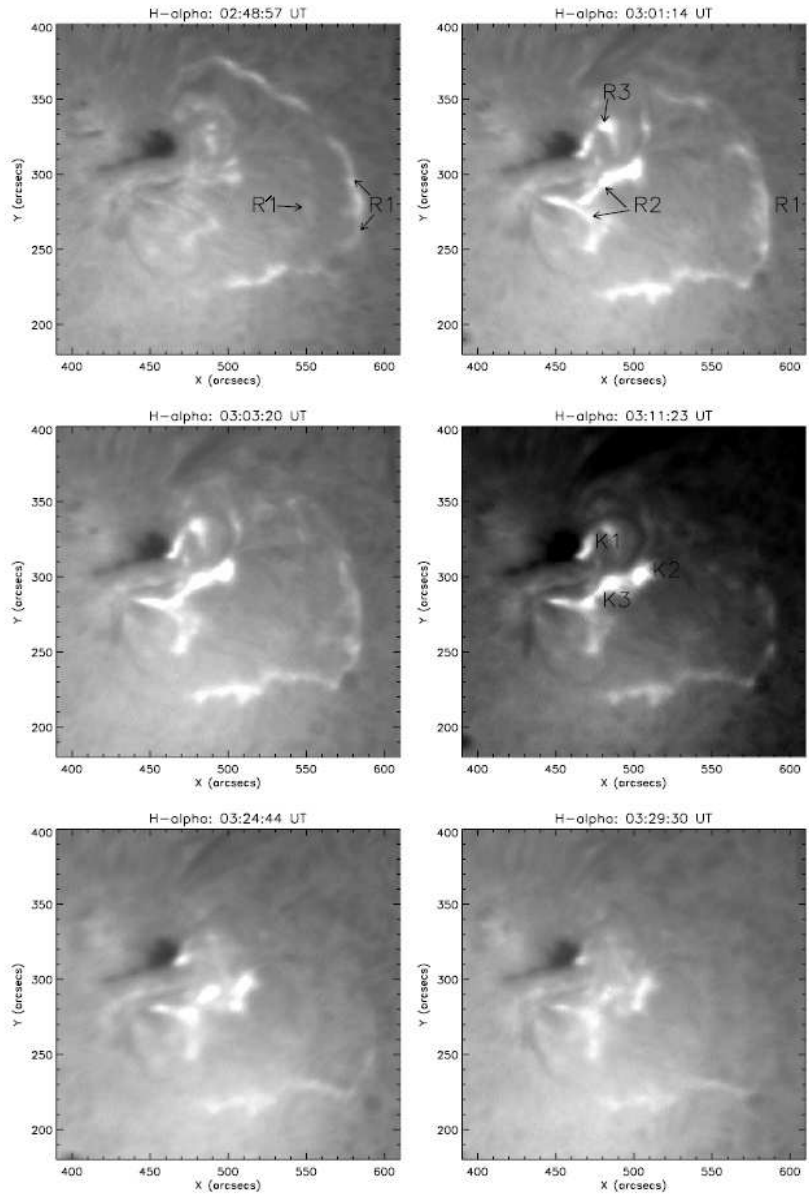}
\caption{Evolution of flare in H$\alpha$ on 26 November, 2000. The locations of different ribbons and kernels are 
marked by R1, R1', R2, R3 and K1, K2, K3 respectively.}
\label{halpha}
\end{figure}

Figure \ref{halpha} shows the evolution of the M2.2/1F class flare in H$\alpha$.
One difference between our studied event here and other previous flare from this
active region is that, most of the previous observed flares were initiated by filament
eruption. In contrary to that the flare of 26 November, 2000 was not associated with any
filament eruption. However, this flare looks homologous to the other flares of this active region.
Similar to other homologous flare, this flare was also associated with
partial halo CME. The description about the associated CME is discussed in subsection 3.3.

\begin{figure}[t]
\centering
\hspace*{-2.0cm}
\includegraphics[width=1.2\textwidth, angle=-90,clip=]{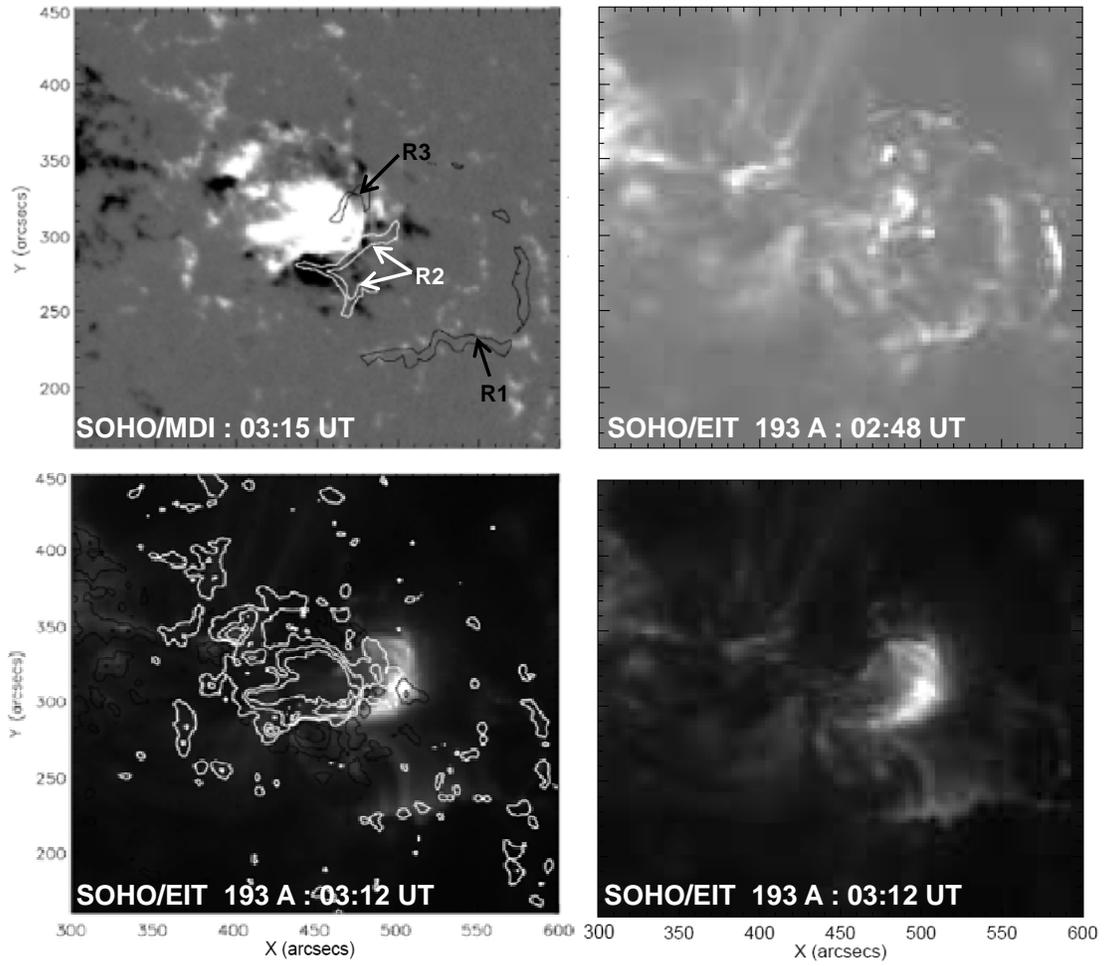}
\caption{Location of H$\alpha$ ribbons on MDI magnetic field(top, left), SOHO/EIT 193 \AA\ image at flare 
onset (top, right), SOHO/EIT 193 \AA\ image overlaid by MDI contours (bottom, left), and peak phase 
of flare in SOHO/EIT 193 \AA\ (bottom, right).}
\label{eit}
\end{figure}

The flare was initiated with arc--shaped ribbon R1 around 02:48 UT. The ribbon R1 has two parts 
eastern ( R1$'$ ) and western ( R1 ). The
eastern fainter ribbon R1$'$ faded quickly. We track the evolution of western ribbon
R1 and found, initially the north portion of ribbon was moving towards west with a
speed of $\sim$ 20 km s$^{-1}$. As the time progress the same part of the ribbon shows
contraction. The north part of the ribbon R1 faded; while as the south part was bright till the end of
the main flare. This ribbon was located in the positive polarity region (see figure \ref{eit}).
Such arc--shaped type ribbons were observed in all the flares produced by 
this active region \citep{wang02, takashi04}. \citet{wang02} interpret these arc--shaped ribbons
as a results of the interaction between erupting flux rope and the overlaying loops.
\citet{wang05} also found the expansion-to-contraction motion in the outer ribbon of 
25 November, 2000 flare event 
from the same active region.  As proposed by \citet{wang05}, this contraction  motion can be interpreted as a 
result of  the falling back of part of erupting flux rope. Such arc--shaped ribbons was also reported in \citet{Chandra09}. 

\begin{figure}
\centering
\hbox{
\hspace*{-0.5cm}
\includegraphics[width=1.0\textwidth,clip=]{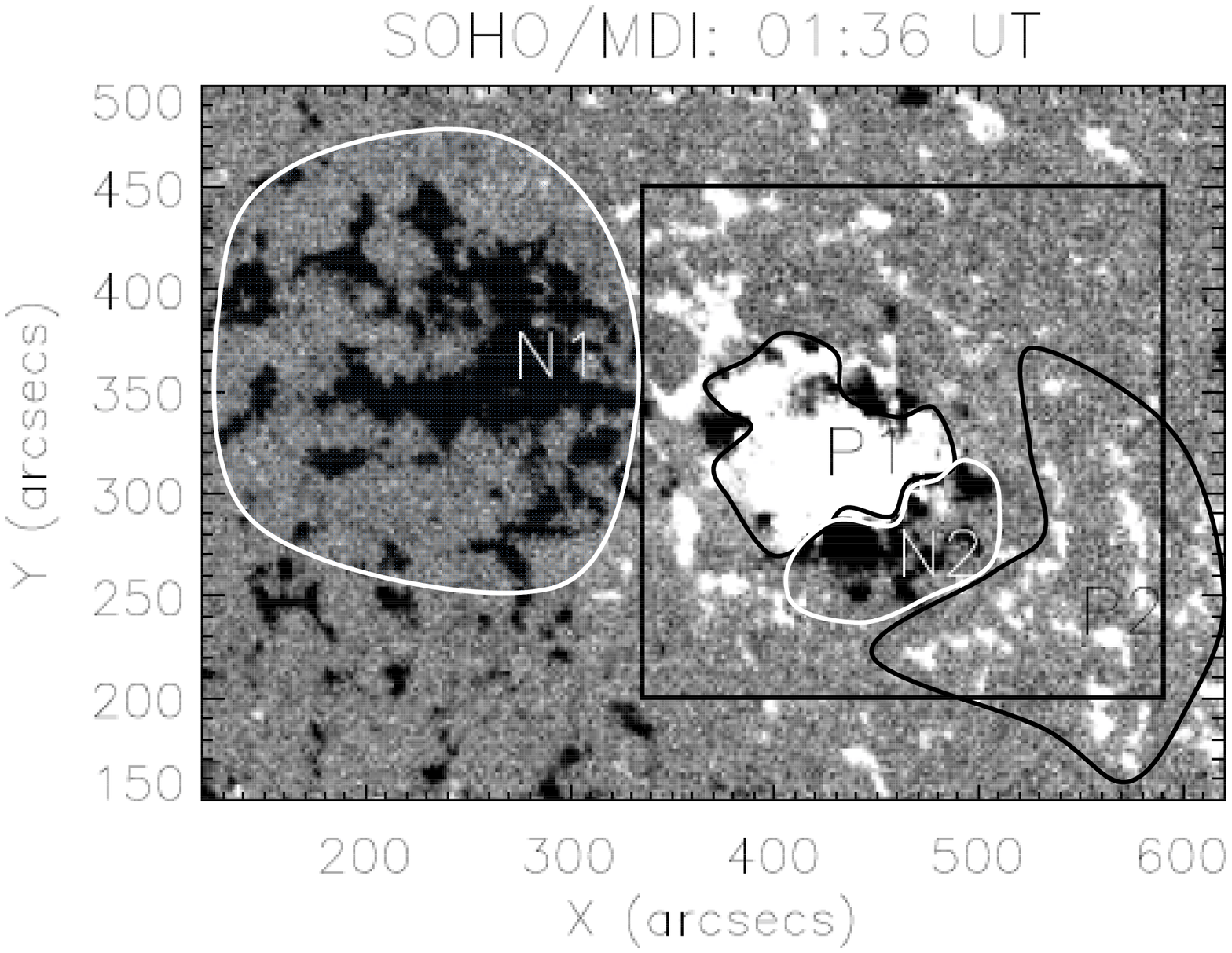}
}

\hbox{
\includegraphics[width=0.5\textwidth,viewport=  10 60 410 420, clip=]{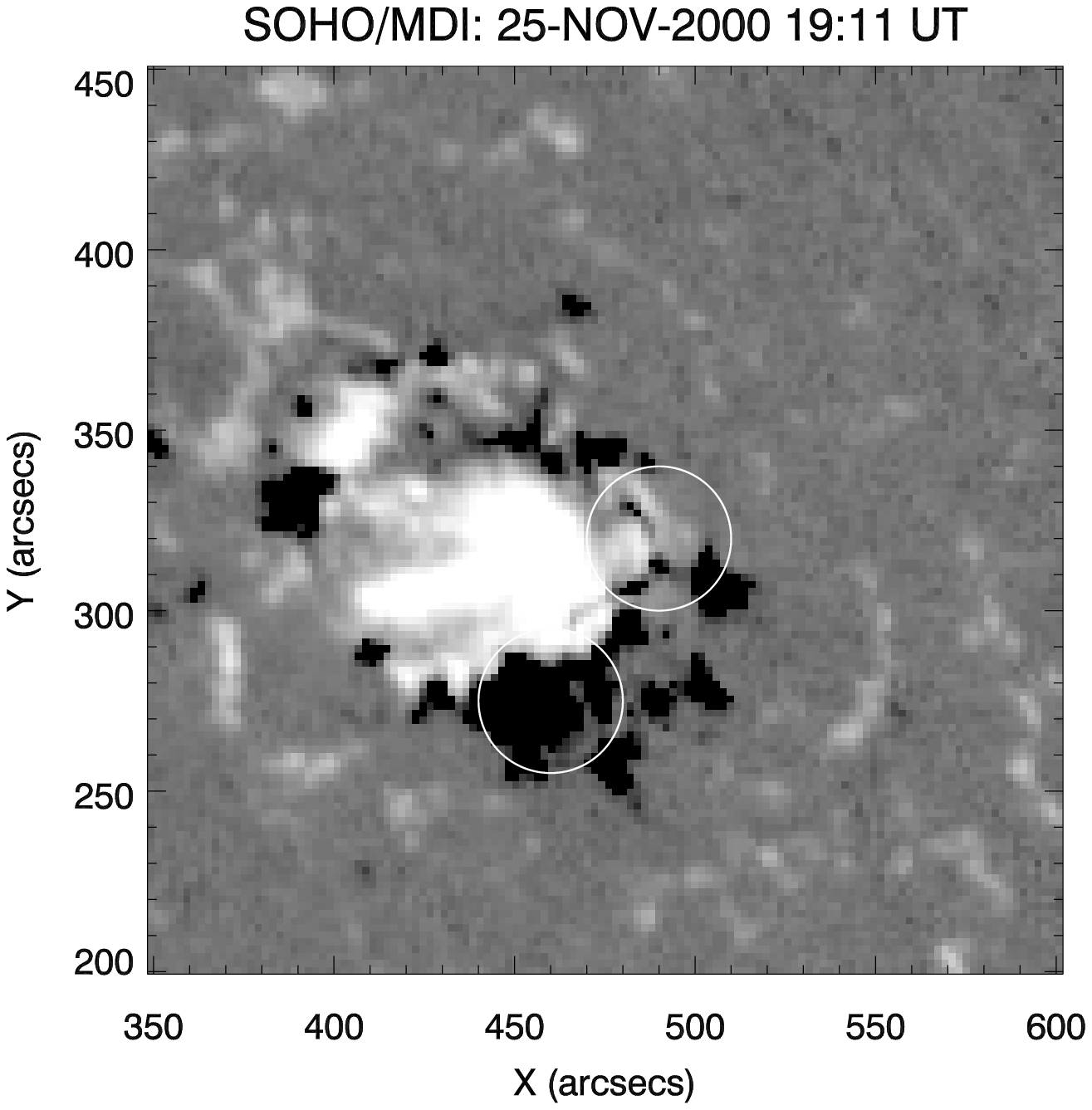}
\includegraphics[width=0.5\textwidth,viewport=  10 60 410 420, clip=]{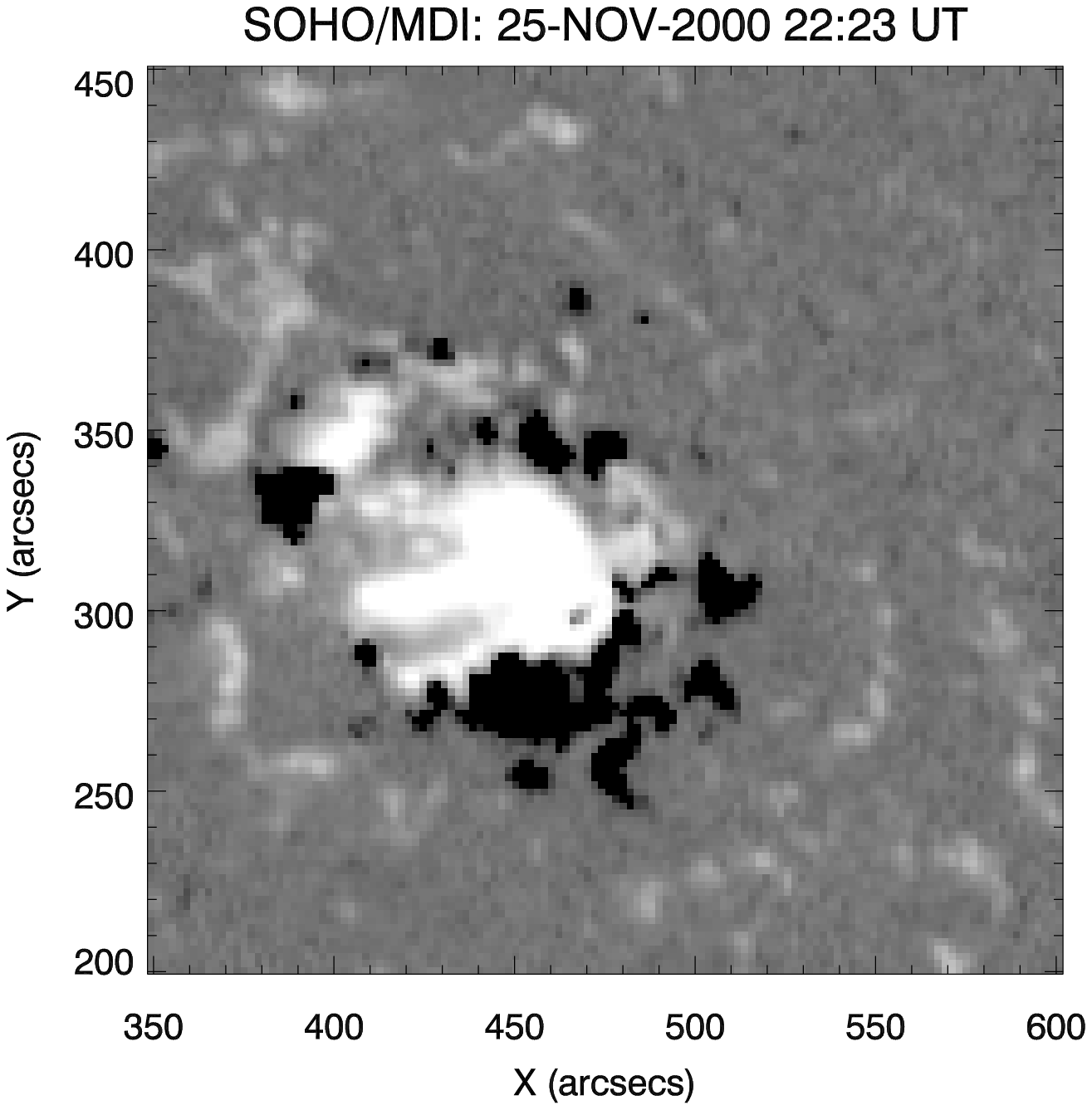}
}
\hbox{
\includegraphics[width=0.5\textwidth,viewport=  0 20 410 420, clip=]{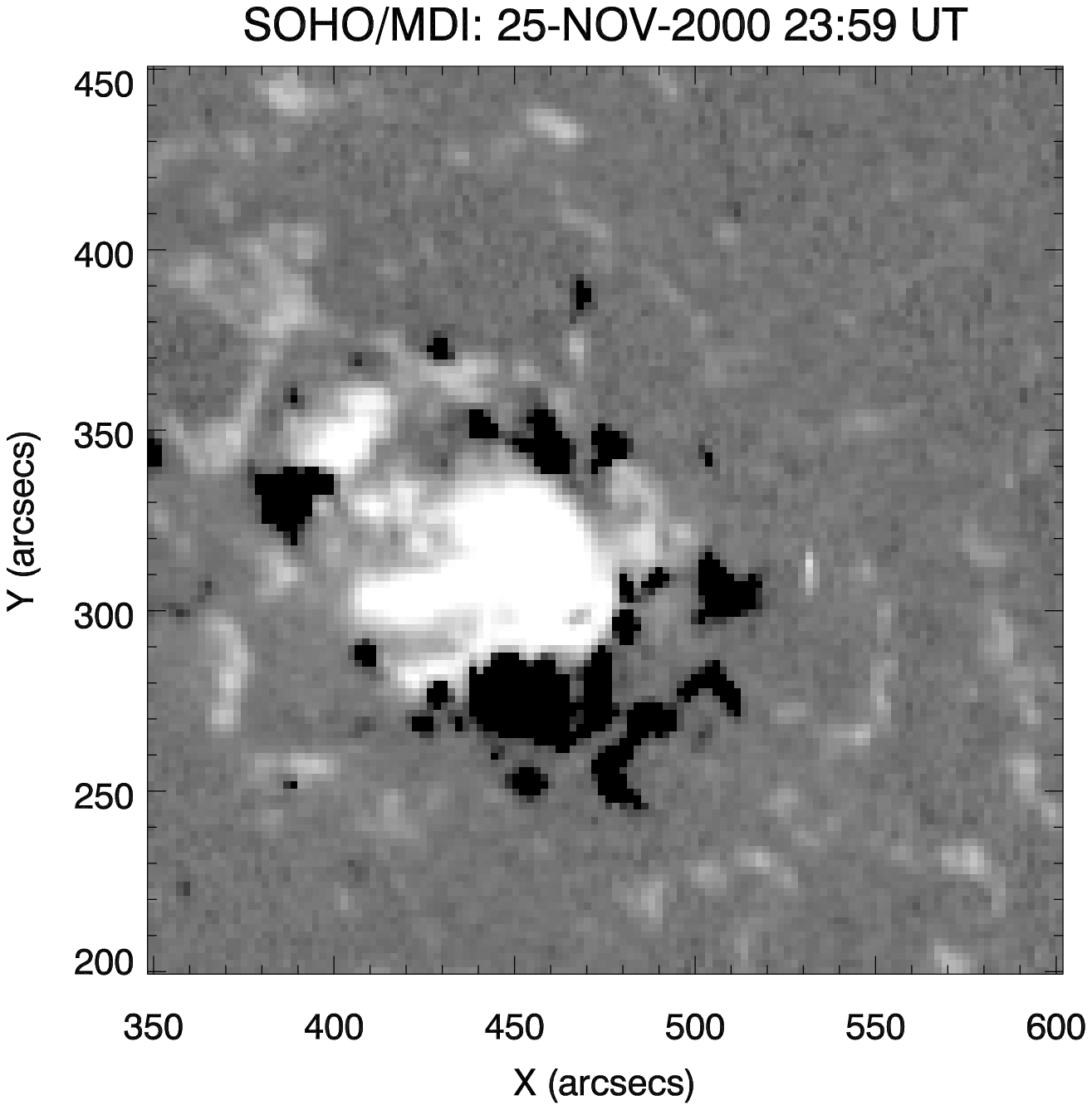}
\includegraphics[width=0.5\textwidth,viewport=  0 20 410 420, clip=]{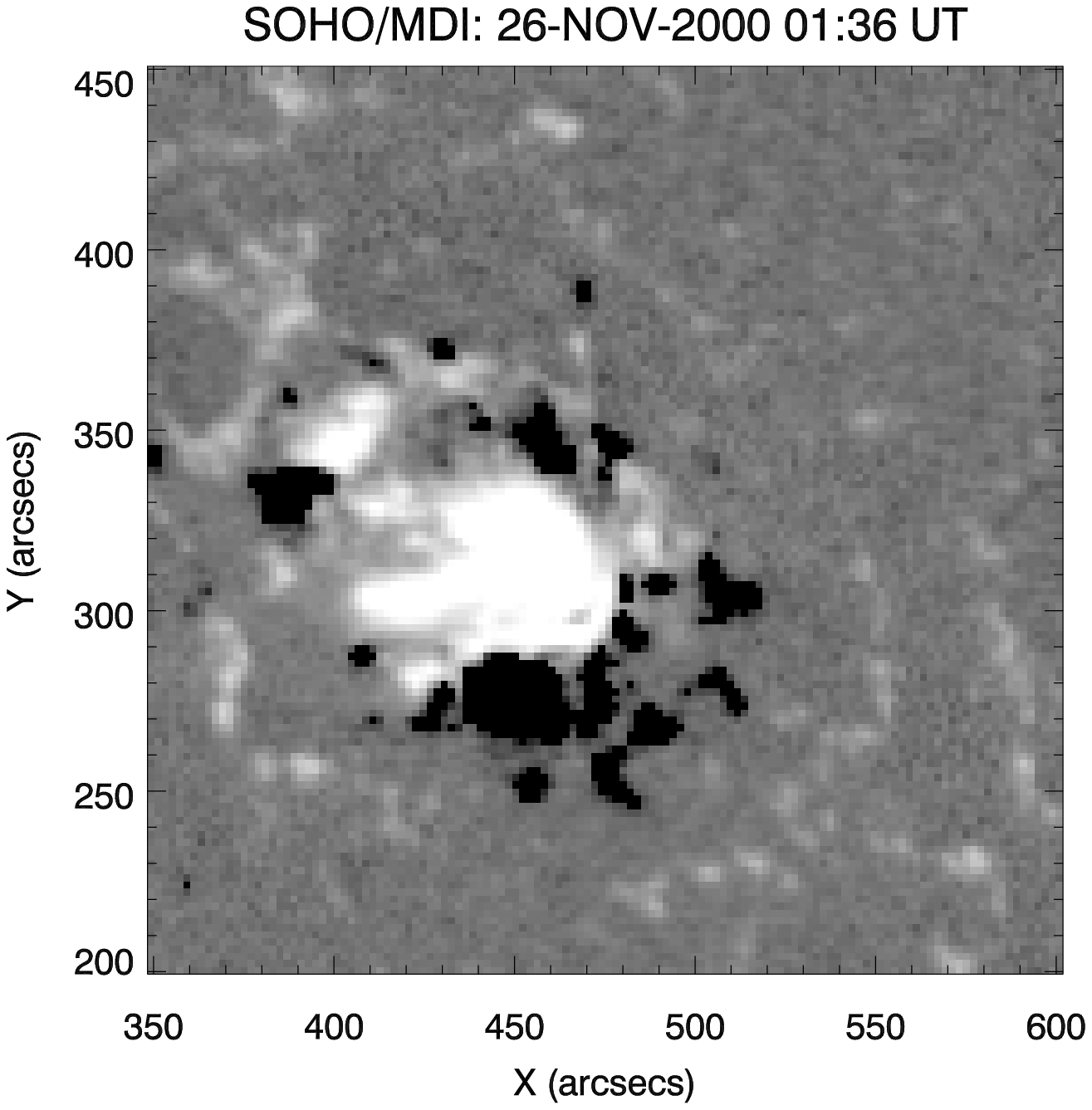}
}
\caption{Top: Full view of active region, The different quadrupole negative/positive polarities are represented by
N1,N2/P1,P2 respectively. The square in the figure represents the field-of-view enlarged in middle and bottom panels. Bottom and bottom panel: Evolution of SOHO/MDI line-of-sight magnetic field before the flare onset. The circles in 
the first image refers to the location, where magnetic flux cancellation occurs.}
\label{magnetic}
\end{figure}

The first appearance of arc--shaped ribbon R1 indicates weak reconnection first and as a result of this the 
ribbon R1 appears. After the weak reconnection, it might weakend the magnetic tension. Afterwards
the main flare occur. This could be a signature of the magnetic breakout  model
as discussed in the introduction section. For the magnetic breakout trigger mechanism, the 
quadrupolar configuration is a necessary condition. If we look at our active region, it has 
also quadrupolar structure (see  figure \ref{magnetic}). 

According to magnetic breakout model proposed by \citet{Antiochos99}, the reconnection start to occur 
above the erupting arcade and not much energy is released during this reconnection.
The reason is that most of the 
free energy is stored in the low laying arcade. Due to the weak reconnection, we might observed the arc--shaped 
outer ribbon in one of the positive polarity `P2' of quadrupolar magnetic configuration 
(see figure \ref{magnetic}, upper panel). The appearance of outer ribbon before the main flare is an 
important implication of the breakout model. Hence, we believe that the quadrupolar magnetic configuration 
together with the first appearances of arc--shaped ribbon R1 strengthen our idea that the flare was 
triggered by magnetic breakout mechanism.

Quickly after the appearance of arc--shaped ribbon R1, main flare ribbons R2
and R3 start to brighten. The main flare ribbons can be seen up to 03:30 UT. As the time
progresses these main flare ribbons show the sign of separation as excepted in CSHKP flare model. To locate 
the location of polarity of flare ribbons in Figure \ref{eit} (top, left panel), the 
contours of flare ribbons are plotted over magnetogram. 
The overplot shows that the main flare ribbon R2 and R3 are located in the positive and 
negative polarities respectively.

Figure \ref{eit} also presented (top right, bottom (left,right)) the SOHO/EIT \mbox{195 \AA} images at 
flare onset and during the flare maximum phase. Figure \ref{eit} (bottom, left) shows the SOHO/EIT
image overlaid by MDI magnetic field contours. The black and white contours represent the negative and
positive polarity respectively. In the bottom, right image, we can see the loops joining the 
ribbon R1 R2 and R2 R3. These loops indicate the connectivity after the flare.
Evolution of main flare ribbons R1 and R2 shows several kernels inside the ribbons. The temporal
evolution of some of the selected kernels is presented in next subsection.

\subsection{Temporal Evolutions}

To study the temporal association of observed flare among different wavelengths, in this
section, we have  presented the intensity evolution of flare as a function of time
observed in H$\alpha$, soft X-rays, hard X-rays, and radio wavelengths. 
As mentioned in previous subsection, during the evolution of the flare in H$\alpha$, we can see 
several kernels. 
we have selected three kernels, as shown in Figure \ref{halpha} at 3:11:25
UT. We name them K1, K2 and K3 respectively. For the computation of flare
kernel intensity, we have created a box of 10 $\times$10 pixels around the kernels and calculate the average
intensity inside it. Afterwards the average intensity is normalized by the
background intensity. For the background intensity we have selected a box of
100$\times$100 pixels in the quiet region. Among the selected kernels K1, and K3 are
strong. Therefore for the temporal evolution, we have plotted only the kernel K1
and K3.

\begin{figure}
\centering
\hspace*{-2cm}
\includegraphics[width=0.8\textwidth, clip=]{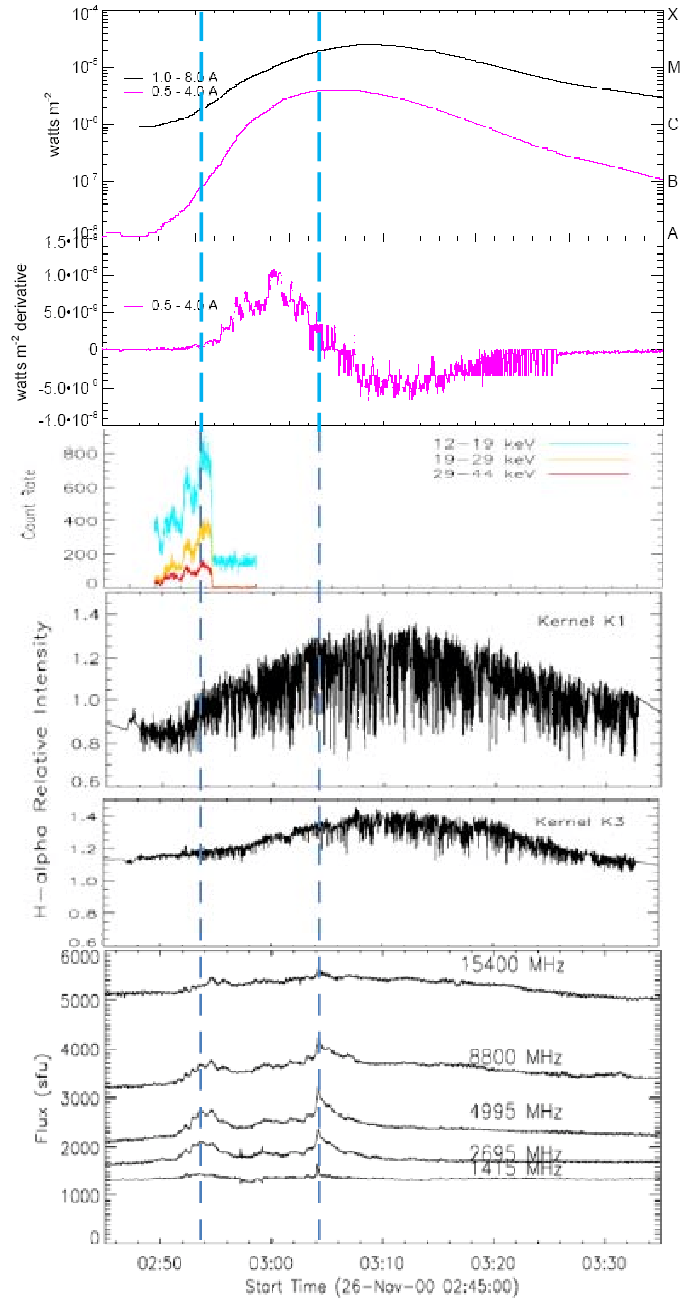}
\caption{Temporal evolution of flare in X-rays, GOES time derivative, H$\alpha$ (flare kernels), and radio at different frequencies.}
\label{temporal}
\end{figure}

In the panel I, II and III 
of Figure \ref {temporal}, we have plotted the GOES X-ray at two wavelengths, GOES X-ray time derivative 
and HXRS hard X-rays, data respectively. The H$\alpha$ kernels are displayed in IV, V of the figure,
whileas VI panel presents the radio observations at frequencies 1415 MHz, 2695
MHz, 4995 MHz, 8800 MHz and 15400 observed from Learmonth Solar
Observatory, Learmonth, Australia.

The impulsive phase of flare (between 02:50 to 02:58 UT) was observed by HXRS at different energy channels. 
The impulsive phase of X-ray peaks around 02:54 UT.
Unfortunately, the HXRS X-ray observations was not available during the peak phase of flare.
According to Neupart effect the time derivative of soft X-ray should correlate with the Hard X-rays.
Therefore, to fill the gap of HXRS observation, We have taken GOES time derivative as a proxy of Hard X-rays.
The time derivative of the GOES X-rays is shown in figure \ref {temporal} (second panel). 
We have noticed the GOES time derivative is peaked earlier ($\sim$ 4 min) then the GOES soft X-ray 
and H$\alpha$. This suggests that the acceleration of non-thermal electrons was involved in the heating 
of lower chromosphere.

Looking at the radio profiles, we noticed two peak: the
first peak was associated with HXRS peaks, whereas the second peak was
associated with main flare phase observed in H$\alpha$ and GOES SXR. The temporal
correlation between the first radio peaks and HXR peaks indicate that the same
population of electrons are responsible for the HXR and radio emissions. 

If we compare the onset of X-rays, radio flux and the appearance of flare ribbon R1. 
They are temporally associated. This indicate that the before the main phase of flare, there is a 
weak reconnection, which triggers the main flare. This is the evidence of flare trigger due to 
magnetic breakout mechanism. The description about the ribbon R1 is discussed in subsection 3.1.

\subsection{CME Observations}

\begin{figure}
\centering
\hspace*{-1cm}
\includegraphics[width=1.2\textwidth, clip=]{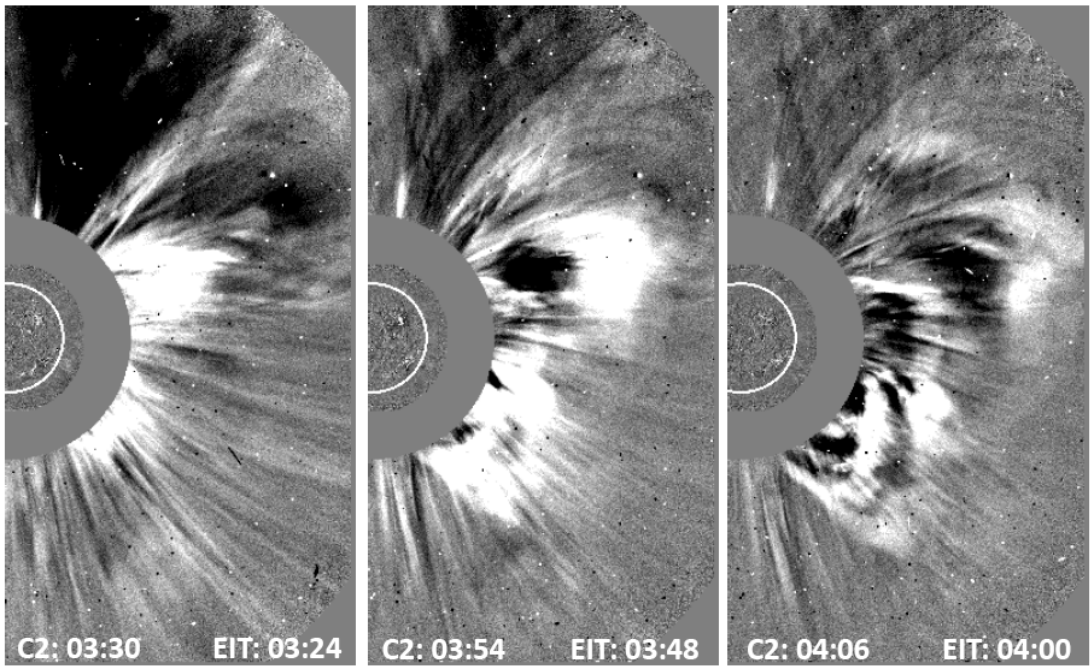}
\caption{Evolution of associated CME observed by LASCO C2 on 26 November, 2000.}
\label{cme}
\end{figure}

The solar flare observed on 2000 November 26 at 02:27 UT was accompanied
by partial CME (width $\sim$ 259 deg) as observed by the LASCO instrument. The CME appear first
in LASCO C2 field-of-view at 03:30 UT in the north-west direction. The evolution
of LASCO C2 CME is displayed in Figure \ref{cme}. The linear velocity of CMEs is 495 km s$^{-1}$. 
The acceleration corresponding to CMEs was -22.9 ms$^{-2}$.

\section{Conclusions}

In this study, we have presented  the high cadence CCD
observations of an M2.2 class solar flare in H$\alpha$ emissions on 2000 November, 26
from NOAA AR 9236. The ground based H$\alpha$ was combined with radio observations and with various space 
borne instruments (SOHO, HXRS, GOES).

The flare started with long arc-shape outer R1 ribbons. Afterwards the main
flare starts with two ribbons. Initially the outer ribbons start to expand with
average speed of 20 km s$^{-1}$ and lateron it shows contraction. The first
appearance of outer flare ribbons and quadrupolar magnetic configuration of active region  
shows that the flare was initiated by the magnetic breakout mechanism \citep{Antiochos99}.
The appearance of X-ray and radio emission flux before the main flare also confirm that the flare is 
triggered by magnetic breakout mechanism.

The SOHO/MDI observations shows emergence of positive and negative magnetic
polarities around the main positive polarity, which might makes this active
region more flare productive. The cancellation of magnetic flux close to the
flare site before the studied M2.2 class flare could provide the evidence that that
flux cancellation is responsible for the flare.

\begin{center}
Acknowledgments 
\end{center}
We thank the anonymous referee for the valuable comments and suggestions.
The authors are thankful to LASCO/SOHO, HXRS and Learmonth Solar
Observatory, Learmonth, Australia whose data are used in the present investigation. The CME
catalog is generated and maintained at the CDAW Data Center by NASA and The Catholic
University of America in cooperation with the Naval Research Laboratory. SOHO is a project of
international cooperation between ESA and NASA.

\bibliography{references}

\end{document}